\documentclass[3p,times,procedia]{llncs}
\usepackage{amsmath}
\usepackage{mathrsfs}
\usepackage{amssymb}
\usepackage{wrapfig}
\usepackage{graphicx}
\usepackage{tikz}
\usepackage{xcolor}
\usepackage{multirow}
\usepackage{booktabs}
\usepackage{pgfplots}
\usepackage[rightComments=false, beginComment=\,\,\,\,\,\,\,\,\,\,\#~]{algpseudocodex}
\usepackage{adjustbox}
\usetikzlibrary{shapes}

\usepackage{geometry}
\newcommand{\cC}{\mathcal C}

\definecolor{lightblue}{HTML}{87CEFA}

\usepackage[bookmarks=false]{hyperref}
    \hypersetup{colorlinks,
      linkcolor=blue,
      citecolor=blue,
      urlcolor=blue}

\usepackage{amssymb}

\begin{document}
\title{Heuristics Optimization of Boolean Circuits with application in Attribute Based Encryption }
\titlerunning{Heuristics Optimisation for BC applied in ABE}
%
\author{Alexandru Ioni\c t\u a \inst{1, 2} \and
Denis Banu\inst{1} \and
Iulian Oleniuc\inst{1}}
\authorrunning{F. Author et al.}
%
\institute{"Alexandru Ioan Cuza" University of Ia\c si, Address, Ia\c si, Romania \and
"Simon Stoilow" Institute of Mathematics of the Romanian Academy, Bucharest, Romania
\email{alexandru.p.ionita@gmail.com}\\
\url{http://profs.info.uaic.ro/~alexandru.ionita} }
\maketitle              

\begin{abstract}
We propose a method of optimizing monotone Boolean circuits by re-writing them in a simpler, equivalent form.  We use in total six heuristics: Hill Climbing, Simulated Annealing, and variations of them, which operate on the representation of the circuit as a logical formula. Our main motivation is to improve performance in Attribute-Based Encryption (ABE) schemes for Boolean circuits. Therefore, we show how our heuristics improve ABE systems for Boolean circuits. Also,  we run tests to evaluate the performance of our heuristics, both as a standalone optimization for Boolean circuits and also inside ABE systems.

\end{abstract}




\section{Introduction}

Modern software systems rely more and more on cloud services for hosting services. These systems bring up a big privacy problem. While using such a service for hosting, including database storage, the Cloud Service Provider usually has access to all the sensitive data, such as client names, addresses, and, depending on the application hosted, possibly medical data or other types of private documents. The obvious solution would be to encrypt the data stored in the cloud. However, this raises the problem of finding expressive encryption systems, in order to obtain a fine-grained access granting mechanism, as we do not wish to offer access to a sensitive document to an unauthorized person.

For example, let's consider a scenario where a healthcare provider's app is hosted in the cloud. There, people could upload medical documents, download medical test results and talk with doctors. In order to grant access only to those who should have it, and to no one else, the old-fashioned way would be to generate a new encryption key for each document and encrypt it with the respective key. However, a document's decryption key should be shared with all the persons who should be able to decrypt it. Having hundreds or thousands of such documents will make this approach infeasible, as each one of the documents will have its own keys.

Here comes in handy Attribute-Based Encryption (ABE), a relatively new encryption system, first introduced in \cite{GPSW2006}. With the ease of ABE, we can encrypt a document under certain attributes (such as ``Cardiovascular", ``Neurological") or even numerically, such as ``Year:2023", ``Priority:3". The decryption keys issued will contain an access structure that operates on such attributes. An example access structure could be: \texttt{(Cardiovascular OR Neurological) AND (Year:2022 OR Year:2021)}.

Multiple documents could be encrypted, each one under its own attribute set. When generating decryption keys, a single decryption key will decrypt many documents, if the attributes in the ciphertexts match the access policy in the key. Thus, we can create a system where we have the ability to grant access control based on descriptive attributes, using a single decryption key for multiple documents.

This type of ABE presented above, where ciphertexts have associated attributes that are matched with the access policies linked to decryption keys, is called \emph{key-policy} ABE. In contrast, there also exists another flavor of ABE, namely \emph{ciphertext-policy} \cite{BeSW2007}, where the access policy is linked to the ciphertext, and the decryption keys have attributes associated with them.

As the complexity of a system increases, so does the complexity of the access structures. Therefore, a challenge for the ABE systems is to find more and more expressive access policies for which we can build efficient ABE systems. For example, the first ABE system \cite{GPSW2006} uses Boolean trees as access structures, where nodes can be \texttt{AND} and \texttt{OR} gates. A more expressive access structure could be a monotone Boolean circuit. Note that not all Boolean circuits can be expressed as access trees, but rather as \emph{directed acyclic graphs}. However, for such access structures, the existing ABE schemes are inefficient \cite{TiDr2014}, \cite{HuGa2017_KP-ABE}, \cite{HuGa2017_CP-ABE} or rely on mathematical primitives for which there is no secure construction known \cite{GGHSW2013} \cite{TiDr2015}.

From this problem we get our motivation: we need to re-write a Boolean circuit in an equivalent form, such that the current ABE systems, like \cite{TiDr2014}, will improve in efficiency.

\vspace{-0.5em}
\subsection{Our Contribution}

\vspace{-0.5em}

The most efficient Attribute-Based Encryption scheme for Boolean circuits from bilinear maps uses a secret sharing technique on Boolean circuits, which results in an exponential expansion in key (or ciphertext) size in the worst-case scenario. Our algorithms are optimizing the access structure -- Boolean circuits -- by finding equivalent circuits, for which the secret sharing is more efficient.

Besides our optimization results, we provide open access to our source code as a library which can be used to optimize Boolean circuits for the \cite{TiDr2014} scheme. Moreover, we also provide an archive with the test cases we used to evaluate the performance of our work, such that subsequent works could be compared to ours using the same datasets.

\vspace{-0.5em}
\section{Related Work}

\vspace{-0.5em}
\subsection{Attribute-Based Encryption for Boolean Circuits}

The first ABE systems were introduced in two flavors, \emph{key-policy} \cite{GPSW2006} and \emph{ciphertext-policy} \cite{BeSW2007}, both of them supporting Boolean trees as access structures.
Then, the problem of finding more expressive access structures arose. Boolean circuits, for example, cover a much larger range of access structures. Unlike a Boolean tree, a Boolean circuit does not limit the fan-out of its gates to one.
Finding an efficient ABE scheme for Boolean circuit access structures is an important open problem in cryptography.
Garg et al.\@ \cite{GGHSW2013} introduced the first ABE system for Boolean circuits. However, their system relies on multi-linear maps and the MDDH assumption, for which there is no known mathematical construction \cite{AlDa2017} \cite{Tipl2018}. Other approaches, such as \cite{TiDr2014} \cite{HuGa2017_CP-ABE}, offer constructions relying on secure and efficient mathematical primitives -- bilinear maps. However, the decryption key could expand exponentially for some circuits.



\subsection{Boolean Circuit Minimization}

The problem of re-writing Boolean circuits in order to obtain a more compact form, with fewer gates is well-known in scientific literature as Boolean Formula Minimization. One of the most well-known algorithms is the Quine-McCluskey Algorithm \cite{Quin1952}, \cite{McCl1956}. However, the problem of Boolean Formula Minimization requires the functions to be given as truth tables. This is impractical for use in an ABE scenario since the Boolean truth table is exponentially larger than the access structure. In ABE we are given an access structure as a Boolean Circuit, and we need to minimize it without computing the truth table.
Another algorithm for Logic optimization of Boolean circuits is Espresso \cite{BHHNS1982}\cite{brayton1984logic}. The main advantage of Espresso is that it uses various heuristics to minimize the circuits, which makes it far more efficient than Quine-McCluskey and allows it to run on higher inputs. It operates on multiple-valued and multiple-output Boolean functions described by min-terms, where each element is assigned a state -- ``True", ``False", or ``Don't care".
Other newer Boolean Minimizers have been proposed, following similar ideas with \cite{SaTC2003}\cite{CoMF1993}. However, all these approaches differ from our scope, since their optimization of Boolean circuits is not the same as the one required for ABE systems. We give more details on this in Section 4.

\subsection{Heuristic Optimizations in Cryptography}
\label{heur_crypt}
The problem of optimizing cryptographic schemes for obtaining better time performance has been approached before. However, most of the time, the optimization is very particular in order to be compatible with a specific cryptosystem. For example, \cite{SaSi2015} presents a survey on such optimizations for cryptographic systems with applicability in Wireless Sensor Networks. The heuristics compared include genetic algorithms and nature-inspired methods such as "Particle Swarm" or "Ant Colony". From our knowledge, there is no previous work on optimizing the decryption phase of ABE using heuristic approaches. The only work related to ABE systems is a very recent work \cite{PiVG2022} that applies a heuristic optimization to ABE for conversion between different types of underlying bilinear maps primitives. This is however a very different type of optimization, both in means and scope, compared to ours.

\section{Prerequisites}

When referring to a logical formula, we distinguish between variables and literals: Variables are symbols that can take values $1$ or $0$, while literals represent  atomic logical formulas (a variable or its negation). Therefore, if a variable appears multiple times in a formula, it is counted each time as a distinct literal.
For example, the formula $A \land B \lor A \land C$ has $3$ variables and $4$ literals.

A Boolean circuit is a Directed Acyclic Graph over a set of input wires, concluding to a single output wire, with internal nodes representing logical gates of type \texttt{AND}, \texttt{OR} or \texttt{NOT}. These gates may have fan-out greater than $1$. A \textit{monotone} Boolean circuit is a circuit without negation gates. A Boolean tree is a Boolean circuit where each node has a fan-out of a maximum of one.

\paragraph{Access Structures \cite{Beim2011}}

Let $\{p_1, \ldots, p_n\}$ be a set of parties. A collection $A \subseteq 2^{\{p_1, \ldots, p_n\}}$ is \textit{monotone} if $(B \in A \land B \subseteq C) \to C \in A$. An access structure is a monotone collection $A \subseteq 2^{\{p_1, \ldots, p_n\}}$ of non-empty subsets of $\{p_1, \ldots, p_n\}$. Sets in $A$ are called \textit{authorized}, while sets not in $A$ are called \textit{unauthorized}.

\subsection{KP-ABE Model}

Key-Policy Attribute-Based Encryption scheme, as first described in \cite{GPSW2006}, consists of four algorithms:

\begin{description}
    \item[setup($\lambda$)] A randomized algorithm that takes as input the implicit security parameter $\lambda$ and returns the public and secret keys ($\mathit{MPK}$ and $\mathit{MSK}$).

    \item[encrypt($m, A, \mathit{MPK}$)] A probabilistic algorithm that encrypts a message $m$ under a set of attributes $A$ with the public key $\mathit{MPK}$, and outputs the ciphertext $E$.

\medskip
    \item[keygen($\cC, \mathit{MPK}, \mathit{MSK}$)] This algorithm receives an access structure $\cC$, public and master keys $\mathit{MPK}$ and $\mathit{MSK}$, and outputs the corresponding decryption keys $\mathit{DK}$.

\medskip
    \item[decrypt($E, \mathit{DK}, \mathit{MPK}$)] Given the ciphertext $E$ and the decryption keys $\mathit{DK}$, the algorithm decrypts the ciphertext and outputs the original message.
\end{description}

\subsection{High-Level Description of Secret Sharing in KP-ABE for Boolean Circuits from \cite{TiDr2014}}

The \emph{key-policy} ABE for Boolean circuits scheme from \cite{TiDr2014} uses bilinear maps as key components to the construction. Therefore, the running time of the four algorithms is strictly related to the number of pairing operations that are computed, since these are by far the most expensive ones. Therefore, our goal will be to minimize the number of pairings that are computed. Unfortunately, due to space limitations, we are not able to provide more details about the bilinear maps as mathematical primitives. However, they are not needed in order to understand our work, but rather just to acknowledge the fact that the pairing operations resulting from the bilinear maps are the most expensive in these ABE systems.

The $\mathit{keygen}$ algorithm uses the Boolean circuit as an access structure in order to generate decryption keys. On the top of the Boolean circuit we will have an output node (we will refer to it as $O_\mathcal{C}$), and each of the input nodes ($\mathit{In}_i$, where $i=\overline{1, n}$) of the circuit will have an attribute (labeled from $1$ to $n$) attached to it. Then, a secret sharing technique will be applied to the circuit top-down, starting from $O_\mathcal{C}$ and ending in the input nodes. Each input node $\mathit{In}_i$ will have some values associated with it. For each of these values, in the decryption phase, we must compute a \textit{pairing} operation.

Due to the construction of the secret sharing technique from \cite{TiDr2014}, the number of shares each attribute will receive in the end is equal to the number of paths from that input node (associated with the attribute) to the output node of the circuit. Therefore, the total number of pairings that will be executed in the decryption phase is equal to the total number of paths from the input nodes to the output node. Therefore, our goal is to find equivalent forms of the circuit such that the total number of paths is minimized.

We will define a cost function $c$ for a circuit $\cC$: $c(\cC)$ should compute the number of shares the secret sharing technique in ABE is producing on $\cC$. This function can also be computed from the logical formula of the circuit: the number of literals in the formula represents the number of paths from the top of the circuit to the bottom since each literal corresponds to a leaf in the Abstract Syntax Tree of the logical formula.

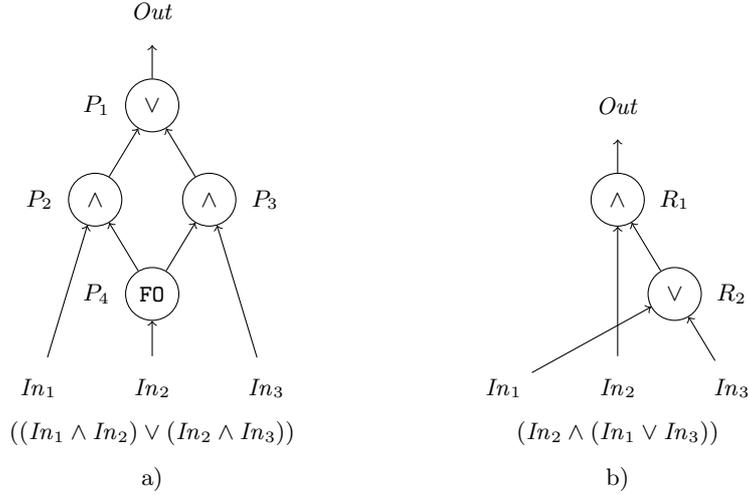
\begin{figure}
\begin{center}
\begin{tikzpicture}[
    every node/.append style = {
        draw,
        shape=circle,
        minimum size=.7cm
    }
]
    \node[draw=none, minimum size=0] (L0) at (1.5, 5) {$\mathit{Out}$};
    \node[draw=none, minimum size=0] (L1) at (0, 0) {$\mathit{In}_1$};
    \node[draw=none, minimum size=0] (L2) at (1.5, 0) {$\mathit{In}_2$};
    \node[draw=none, minimum size=0] (L3) at (3, 0) {$\mathit{In}_3$};
    \node[label=left:$P_1$] (P1) at (1.5, 3.75) {$\lor$};
    \node[label=left:$P_2$] (P2) at (.75, 2.5) {$\land$};
    \node[label=right:$P_3$] (P3) at (2.25, 2.5) {$\land$};
    \node[label=left:$P_4$] (P4) at (1.5, 1.25) {$\small \texttt{FO}$};
    \draw[<-] (L0) -- (P1);
    \draw[<-] (P1) -- (P2);
    \draw[<-] (P1) -- (P3);
    \draw[<-] (P2) -- (P4);
    \draw[<-] (P3) -- (P4);
    \draw[<-] (P2) -- (L1);
    \draw[<-] (P4) -- (L2);
    \draw[<-] (P3) -- (L3);
    \node[draw=none, minimum size=0, shape=rectangle] (formula) at (1.5, -0.6) {$((\mathit{In}_1 \land \mathit{In}_2) \lor (\mathit{In}_2 \land \mathit{In}_3))$};
    \node[draw=none, minimum size=0, shape=rectangle] (label) at (1.5, -1.2) {a)};
\end{tikzpicture}
\hspace{2cm}
\begin{tikzpicture}[
    every node/.append style = {
        draw,
        shape=circle,
        minimum size=.7cm
    }
]
    \node[draw=none, minimum size=0] (L0) at (1.5, 3.75) {$\mathit{Out}$};
    \node[draw=none, minimum size=0] (L1) at (0, 0) {$\mathit{In}_1$};
    \node[draw=none, minimum size=0] (L2) at (1.5, 0) {$\mathit{In}_2$};
    \node[draw=none, minimum size=0] (L3) at (3, 0) {$\mathit{In}_3$};
    \node[label=right:$R_1$] (R1) at (1.5, 2.5) {$\land$};
    \node[label=right:$R_2$] (R2) at (2.25, 1.25) {$\lor$};
    \draw[<-] (L0) -- (R1);
    \draw[<-] (R1) -- (R2);
    \draw[<-] (R1) -- (L2);
    \draw[<-] (R2) -- (L1);
    \draw[<-] (R2) -- (L3);
    \node[draw=none, minimum size=0, shape=rectangle] (formula) at (1.5, -0.6) {$(\mathit{In}_2 \land (\mathit{In}_1 \lor \mathit{In}_3))$};
    \node[draw=none, minimum size=0, shape=rectangle] (label) at (1.5, -1.2) {b)};
\end{tikzpicture}
\caption{Two equivalent Boolean circuits, alongside their equivalent logical formulas} \label{fig_abe}
\end{center}
\end{figure}

One good example of two equivalent Boolean circuits is depicted in Fig.\@ \ref{fig_abe}. The first circuit (Fig.\@ \ref{fig_abe}a) leads in the secret sharing scheme from {\c{T}}iplea-Dr{\u{a}}gan scheme \cite{TiDr2014} to a total number of 4 shares: one for $\mathit{In}_1$, two for $\mathit{In}_2$ and one for $\mathit{In}_3$. However, the equivalent circuit from Fig.\@ \ref{fig_abe}b leads only to 3 shares. Therefore, the decryption time will be roughly 25\% smaller if we use the second circuit.

\section{Boolean Circuit Minimization for ABE}

Since the problem of Boolean Minimization is well-studied, the first obvious choice while would be to try to use an existing algorithm. However, the existing algorithms optimize the circuit for constructing Programmable Logic Arrays (PLAs). Therefore, the input and output circuits will be in a format similar to DNF, which allows the easy construction of a PLA. Even if we convert the Boolean circuits from ABE's input to a logic minimizer such as Espresso, then we would have to also process the output. The DNF format in which these minimizers output the formula is an inefficient form for a circuit, w.r.t.\@ the secret sharing scheme used in ABE. The DNF is the most uncompressed form a circuit can have. The logic minimizers are trying to optimize the number of gates used, while we want to minimize the number of literals in the logic expression associated with the circuit, or the number of paths from the output to the input nodes.







\subsection{The Approach}

We thought about how we can obtain, in general, equivalent logical formulas, starting from some given expression. We have observed three operations that we may apply to a monotone (without negations) logical formula to obtain equivalent ones. We will refer to these operations as \emph{factorization}, \emph{defactorization} and \emph{absorption}:

\begin{enumerate}
    \item \emph{factorization} -- Using the fact that $\texttt{OR}$ is distributive under $\texttt{AND}$ (and vice-versa), we can search for common factors inside logical formulas. For example, $(A\land B) \lor (A \land C)$ can be factorized into $A\land (B \lor C)$. Note that this operation always obtains a formula with a strictly lower cost, since it reduces the common factor. 
    
    \item \emph{defactorization} -- This is the inverse operation of \emph{factorization}. We choose a conjunction, and we ``split the parenthesis", resulting in a cross-product of the elements involved. For example, $A\land (B\lor C)$ can be defactorized into $(A\land B) \lor (A \land C)$. Note that this operation gives us an equivalent expression, but with a strictly higher cost.

    \item \emph{absorption} is the operation of eliminating $1$s after the factorization process.
    For example, $A \lor (A \land B)$ can be factorized into something like $A \land (1 \lor B)$. However, the $B$ term is actually ``shadowed" by $1$. Therefore, we can replace the initial formula with $A$, ignoring the term $B$. In our implementation, we have embedded the \emph{absorption} in the \emph{factorization} procedure. Therefore, throughout the rest of the paper, we will only refer to the first two operations -- \emph{factorization} and \emph{defactorization}.
\end{enumerate}

\subsection{Implementation of Operations}

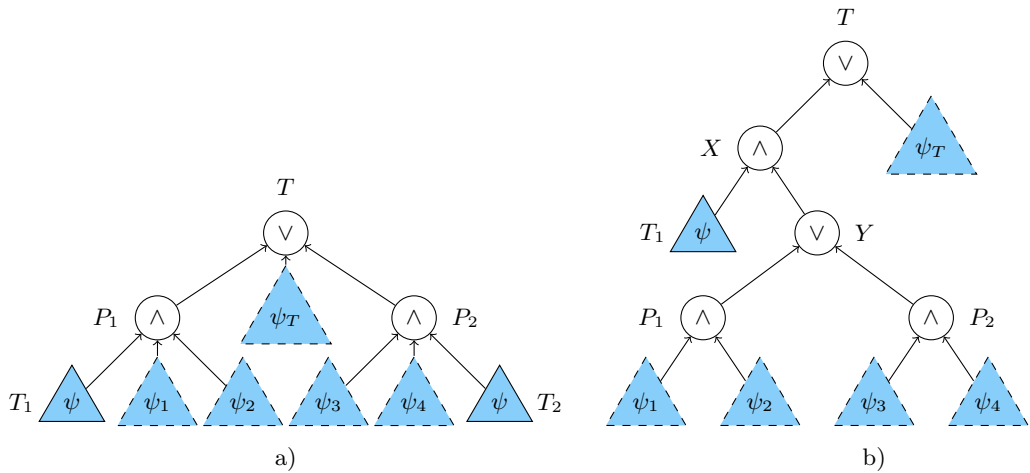
\begin{figure}
\begin{center}
\begin{tikzpicture}[
    every node/.append style = {
        draw,
        shape=circle
    },
    scale=.75
]
    \node[label=left:$T_1$, regular polygon, regular polygon sides=3, inner sep=1pt, minimum size=1pt, fill=lightblue] (T1) at (-.75, 0) {$\psi$};
    \node[regular polygon, regular polygon sides=3, inner sep=1pt, minimum size=1pt, fill=lightblue, dashed] (PSI1) at (.75, 0) {$\psi_1$};
    \node[regular polygon, regular polygon sides=3, inner sep=1pt, minimum size=1pt, fill=lightblue, dashed] (PSI2) at (2.25, 0) {$\psi_2$};
    \node[regular polygon, regular polygon sides=3, inner sep=1pt, minimum size=1pt, fill=lightblue, dashed] (PSI3) at (3.75, 0) {$\psi_3$};
    \node[regular polygon, regular polygon sides=3, inner sep=1pt, minimum size=1pt, fill=lightblue, dashed] (PSI4) at (5.25, 0) {$\psi_4$};
    \node[label=right:$T_2$, regular polygon, regular polygon sides=3, inner sep=1pt, minimum size=1pt, fill=lightblue] (T2) at (6.75, 0) {$\psi$};
    \node[label=left:$P_1$] (P1) at (.75, 1.5) {$\land$};
    \node[label=right:$P_2$] (P2) at (5.25, 1.5) {$\land$};
    \node[regular polygon, regular polygon sides=3, inner sep=1pt, minimum size=1pt, fill=lightblue, dashed] (PSIT) at (3, 1.5) {$\psi_T$};
    \node[label=above:$T$] (T) at (3, 3) {$\lor$};
    \draw[<-] (T) -- (P1);
    \draw[<-] (T) -- (P2);
    \draw[<-] (T) -- (PSIT);
    \draw[<-] (P1) -- (T1);
    \draw[<-] (P1) -- (PSI1);
    \draw[<-] (P1) -- (PSI2);
    \draw[<-] (P2) -- (PSI3);
    \draw[<-] (P2) -- (PSI4);
    \draw[<-] (P2) -- (T2);
    \node[draw=none, minimum size=0, shape=rectangle] (label) at (3, -1) {a)};
\end{tikzpicture}
\hspace{.15cm}
\begin{tikzpicture}[
    every node/.append style = {
        draw,
        shape=circle
    },
    scale=.75
]
    \node[label=left:$T_1$, regular polygon, regular polygon sides=3, inner sep=1pt, minimum size=1pt, fill=lightblue] (T1) at (0, 1.5) {$\psi$};
    \node[label=left:$P_1$] (P1) at (0, 0) {$\land$};
    \node[label=right:$P_2$] (P2) at (4, 0) {$\land$};
    \node[label=right:$Y$] (Y) at (2, 1.5) {$\lor$};
    \node[label=left:$X$] (X) at (1, 3) {$\land$};
    \node[regular polygon, regular polygon sides=3, inner sep=1pt, minimum size=1pt, fill=lightblue, dashed] (PSIT) at (4, 3) {$\psi_T$};
    \node[regular polygon, regular polygon sides=3, inner sep=1pt, minimum size=1pt, fill=lightblue, dashed] (PSI1) at (-1, -1.5) {$\psi_1$};
    \node[regular polygon, regular polygon sides=3, inner sep=1pt, minimum size=1pt, fill=lightblue, dashed] (PSI2) at (1, -1.5) {$\psi_2$};
    \node[regular polygon, regular polygon sides=3, inner sep=1pt, minimum size=1pt, fill=lightblue, dashed] (PSI3) at (3, -1.5) {$\psi_3$};
    \node[regular polygon, regular polygon sides=3, inner sep=1pt, minimum size=1pt, fill=lightblue, dashed] (PSI4) at (5, -1.5) {$\psi_4$};
    \node[label=above:$T$] (T) at (2.5, 4.5) {$\lor$};
    \draw[<-] (T) -- (X);
    \draw[<-] (T) -- (PSIT);
    \draw[<-] (X) -- (T1);
    \draw[<-] (X) -- (Y);
    \draw[<-] (Y) -- (P1);
    \draw[<-] (Y) -- (P2);
    \draw[<-] (P1) -- (PSI1);
    \draw[<-] (P1) -- (PSI2);
    \draw[<-] (P2) -- (PSI3);
    \draw[<-] (P2) -- (PSI4);
    \node[draw=none, minimum size=0, shape=rectangle] (label) at (3, -2.5) {b)};
\end{tikzpicture}
\caption{Modification of an AST after a factorization} \label{fig_trees}
\end{center}
\end{figure}

In order to find a common factor inside multiple terms in a Boolean expression, we make use of the Abstract Syntax Tree (AST) associated with it.
Let $\mathcal{T_\varphi}$ be the AST for some formula $\varphi$. For each node $T_i$, we will denote with
$f(T_i) $ the logical formula associated with the subtree rooted in $T_i$, with $\mathit{parent}(T_i)$ the parent node of $T_i$, and with $\mathit{children}(T_i)$ the set of children nodes. Then, in order to get a common factor, we need to get two nodes $T_1$ and $T_2$ such that:
\begin{itemize}
    \item[--] $f(T_1) = f(T_2)$ -- The formula $f(T_1)$ will be the common factor.
    \item[--] $\mathit{parent}(T_1) \neq \mathit{parent}(T_2)$ -- This is more of a consistency check. A well-formed formula should not have in the AST two siblings with the same formula, as one of them is clearly irrelevant.
    \item[--] $\mathit{parent}(\mathit{parent}(T_1)) = \mathit{parent}(\mathit{parent}(T_2))$ -- Nodes $T_1$ and $T_2$ should have a common grandparent in the AST, as shown in Fig.\@ \ref{fig_trees}.
\end{itemize}
After this operation, the overall cost of the circuit will drop by $c(T_2)$, since we eliminate this part of the circuit.

There are some particular cases in the \emph{factorization} and \emph{defactorization} processes, but for the sake of simplicity, we decided to omit these cases here.



\subsection{Hill Climbing}

The factorization operation will always reduce the cost of the formula. It made us think it is suitable to be used in a Hill Climbing algorithm as we always move towards an optimum and at the end we will reach that optimum. However, the found optimum can be a \textit{local} optimum and it means that we depend on the initial form of the formula. Therefore, until there is no possible factorization left, we randomly choose two nodes that can be factorized, and we apply the operation. When there is no factorization possible anymore, it means we have reached a local maximum.

\subsection{Simulated Annealing}

Simulated Annealing, proposed first time in \cite{kirkpatrick1983optimization}, is a probabilistic method for finding the global minimum of a cost function. This method is inspired by the cooling of metals. In general, at each iteration of the algorithm, a new solution is considered. The probability that the solution is accepted varies with the temperature and the solution score. The higher the temperature, the higher the probability of accepting the solution. Similarly the higher the score, the higher probability of accepting the solution.

We have a temperature that starts with a value $t_{\max}$ and drops over time with a cooling rate $c$. The acceptance probability function used is $e^{-\Delta t}$, where $\Delta$ is the difference between the neighbor formula cost and the current formula cost. The neighbor is obtained from the current formula by randomly choosing one operation between factorization and defactorization, and a random place where it can be applied in the circuit. The probability that we used to pick defactorization is $25\%$. Then, if defactorization is chosen, the probability to accept the new neighbor is given by the acceptance function. The algorithm looks as follows:

\vspace{1em}
\begin{algorithmic}[1]
    \State $t \gets t_{\max}$
    \For{$i \in \{1, 2, \ldots, L\}$}
        \State $\mathit{operation} \gets \texttt{defactorization} \textbf{ if } \operatorname{random}(0, 1) < d \textbf{ else } \texttt{factorization}$
        \State choose a neighbor $u$ using $\mathit{operation}$
        \State $\Delta \gets \operatorname{cost}(u) - \mathit{current\_cost}$
        \If{$\mathit{operation} = \texttt{defactorization}$}
            \State accept $u$ with probability $e^{-\Delta t}$
        \Else
            \State accept $u$
        \EndIf
    \EndFor
    \State $t \gets (1 - c) \cdot t$ \Comment{cooling}
    \If{$t > t_{\min}$} \State \textbf{goto} 2
    \EndIf
    \State apply \texttt{factorization} until formula is not improved anymore
\end{algorithmic}
\vspace{1em}

We used the following parameters: $t_{\max}=100$, $t_{\min}=10$, $c=0.1$, $l=25$. The values for $t_{\max}$ and $t_{min}$ were chosen in this way because the cost of each logical formula from the dataset is between $3$ and $100$.

The factorization operation will decrease the cost, while defactorization will increase it. Using the simulated annealing algorithm, at the beginning we have a higher chance to accept defactorization in order to better explore the solution space. During this time, the rate of acceptance for defactorization decreases until we accept only factorizations, in order to improve the final solution as much as possible.

\subsection{Custom Heuristic}

Besides the classical Hill Climbing and Simulated Annealing heuristics we tried to create a new one that is meant to combine both \textit{factorization} and \textit{defactorization} operations but in a simpler way than Simulated Annealing. In this algorithm, we have $k_{\max}$ iterations. At each iteration, we choose either to factorize or defactorize the current formula. At iteration $k$ the probability to choose defactorization is $\frac{k_{\max} - k}{5k_{\max}}$. It means that, at first iteration, we have a 25\% chance to choose defactorize and it slowly decreases until reaching 0\% at the last iteration.

Then we choose the formula with the smallest cost among all $k_{\max}$ iterations. Finally, we apply factorization on this formula until it can't be improved anymore.

\subsection{Iterated Versions}

In the simple Hill Climbing algorithm the solution converges very quickly to a local optimum. But, if at one step we choose to factorize a different pair of nodes, we can end up in a different local optimum, which can have a smaller cost.

In order to find better solutions, we run the Hill Climbing algorithm multiple times and choose the best solution among them. This gives us the opportunity to explore more and finally, we can end up in the \textit{global} optimum. However, there is the possibility that the best equivalent formula (the \textit{global} optimum) can't be obtained from the initial formula only by doing \textit{factorizations}. There are many formulas where Iterated Hill Climbing gives good results but it can't find the global optimum regardless of the number of iterations we give.

For similar reasons, we have also constructed iterated versions of our other heuristics: Simulated Annealing and the Custom Heuristic. Finally, we have six algorithms -- three main ones and three iterated versions of them.

\section{Practical Tests}


\subsection{Dataset Description}

\begin{wraptable}{r}{6.5cm}
\begin{center}
\begin{tabular}{c|c|c}
 & Variable count & Literal count\\ 
\hline
Dataset 1 & $20-25$ & $20-40$\\ 
Dataset 2 & $20$ & $60-90$ \\ 
Dataset 3 & $25-35$ & $160-200$ \\  
Dataset 4 & $20-25$ & $20-40$  
\end{tabular}
\caption{\label{datasets}Dataset parameters}
\end{center}
\end{wraptable}

We have generated four datasets, each of them with some particularities. The first three datasets consist of randomly generated logical formulas. Their numbers of variables and literals respect the values in Table \ref{datasets}. In this case, we tried to simulate real-world scenarios, where, when dealing with access structures, the most usual way of defining an access structure is by enumerating the minimum sets of participants which should have access to decrypt the data. Therefore, we constructed our formulas bottom-up by constructing formulas for minimum authorized sets, and then linking them together with AND or OR nodes.

Moreover, in the generation of all our datasets, we have run a special procedure called ``trim", which ensures that the generated logical formulas do not have obvious design flaws. One such example could be the formula: $(A\land B)\lor (A\land B) \lor (A\land C)$. Here it is obvious that one of $(A \land B)$ could be removed from the expression.

\begin{wrapfigure}{l}{5.25cm}
\centering
\begin{tikzpicture}
%

\node[circle,draw,label=left:,minimum size=.7cm,inner sep=0pt] (a0) at (3, 1) {$A_0$};
\node[circle,draw,label=left:,minimum size=.7cm,inner sep=0pt] (a1) at (1.5, 1) {$A_1$};
\node[circle,draw,label=left:,minimum size=.7cm,inner sep=0pt] (a2) at (0.75, 2) {$A_2$};
\node[circle,draw,label=left:,minimum size=.7cm,inner sep=0pt] (a3) at (0, 3) {$A_3$};
\node[circle,draw,label=left:,minimum size=.7cm,inner sep=0pt] (a4) at (-0.75, 4) {$A_4$};

\node[circle,draw,label=left:,minimum size=.7cm,inner sep=0pt] (gate1) at (2.25, 2) {$\land$};
\node[circle,draw,label=left:,minimum size=.7cm,inner sep=0pt] (gate2) at (1.5, 3) {$\lor$};
\node[circle,draw,label=left:,minimum size=.7cm,inner sep=0pt] (gate3) at (0.75, 4) {$\land$};
\node[circle,draw,label=left:,minimum size=.7cm,inner sep=0pt] (gate4) at (0, 5) {$\lor$};

\path (gate1) edge[<-] (a0) edge[<-] (a1);
\path (gate2) edge[<-] (gate1) edge[<-] (a2);
\path (gate3) edge[<-] (gate2) edge[<-] (a3);
\path (gate4) edge[<-] (gate3) edge[<-] (a4);

\end{tikzpicture}
\caption{Subcircuit for the comparison ``$A \ge 11$" ($11 = 1011_{(2)}$)}
\label{Fig_comp}
\end{wrapfigure}
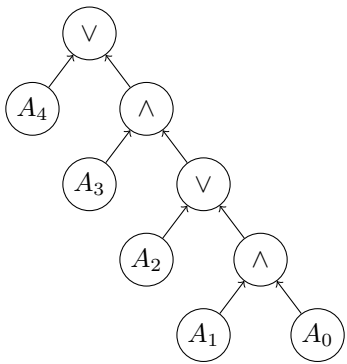
\subsection{Results}
The fourth test is created by taking a concrete example of a complex access policy that can arise in practical systems: comparison queries. There are several works on ABE with access structures supporting such queries \cite{Ioni2022} \cite{BeSW2007} \cite{ZHAYZ2012}. When dealing with numerical numbers as attributes, the access structure may require to allow decryption for values smaller than some threshold value. Therefore, we created the fourth dataset with such cases. In order to perform a comparison query, a numerical attribute $A$ in the range $0$ to $N$ can be split into $\log_2(N)$ smaller attributes, each of them representing a bit in $A$'s binary representation. Then, we can create Boolean circuits which can handle multiple comparison queries. An example of such a Boolean circuit is seen in Fig.\@ \ref{Fig_comp}. There, $A_0, A_1, A_2, A_3, A_4$ represent attributes for the bits of the numeral value of $A$. The attribute $A_i$ is True if and only if $A$ has the $i$th bit set in its binary representation. The Access structures corresponding to such Boolean circuits could be something as \texttt{((Year <= 2022 AND Year >= 2020) OR (Year <= 1990)) AND (Month >= 2 AND Month <= 5)}.\hfill\,

Since our heuristics are probabilistic, we want to compute the expected optimization we will obtain. Therefore, for each formula, we run each algorithm $16$ times and we computed the average of the optimized percentage after each run. Also, we keep track of the maximum optimization obtained over all iterations. This value will be close to the upper bound of the algorithms. In Table \ref{results} we show the \textit{mean optimization} (MO), \textit{best over iterations} (BOI), and \textit{average running time} (ART) for all the formulas in each set. We have run all three algorithms presented above: HC stands for \emph{Hill Climbing}, SA for \emph{Simulated Annealing}, and CH for \emph{Custom Heuristic}. We also have the iterated versions of these algorithms, denoted with IHC, ISA, and ICH, respectively.

We see that Hill Climbing (HC in the table) is by far the fastest, and it also gives decent results for the first three datasets; however they are very low for the one with more practical formulas. Iterated Simulated Annealing (ISA) obtains some very good results on all sets, but its running time is the largest, making it even slower than Iterated Custom Heuristic. The latter beats (or at least ties) all the other algorithms on every set and every metric, and it also has a decently low running time.


Depending on the circumstances, one of the algorithms above could be chosen to optimize the ABE access structure, having a trade-off between running time and optimization. In order to better understand this trade-off, we have made the following experiment:
\begin{enumerate}
    \item Generate 5 Boolean circuit access structures, with 50, 100, 150, 200, and 250 literals each.
    \item For each of these access structures, run the HC, CH, and ISA optimization heuristics presented in this paper.
    \item Construct an ABE system with each of these access structures (20 in total) and run the KeyGeneration  and Decryption (after a previous Encryption) algorithms.
    \item Add the running time of the optimization heuristic to the KeyGeneration algorithm.
    \item Repeat steps 1-4 for 30 times and compute the mean value for each case.
\end{enumerate}

\begin{adjustbox}{tabular=lll,center, caption={(Iterated) Hill Climbing/ Simulated Annealing/ Custom Heuristic results on each dataset}, float=table}
\bgroup
\def\arraystretch{1.3}
\setlength\tabcolsep{0.5em}
\begin{tabular}{ c||c c c|c c c|c c c|c c c }

  &  \multicolumn{3}{c|}{Dataset 1} & \multicolumn{3}{c|}{Dataset 2} &
\multicolumn{3}{c|}{Dataset 3} &
\multicolumn{3}{c}{Dataset 4}\\
  &  MO & BOI & ART & MO & BOI & ART & MO & BOI & ART & MO & BOI & ART \\
 \hline
 \hline
 HC & 15.0 \% & 16.3 \% & 0.00 s & 35.1 \% & 41.9 \% & 0.00 s & 42.6 \% & 56.5 \% & 0.01 s & 4.8 \% & 7.2 \% & 0.00 s\\
\hline
IHC & 16.3 \% & 16.3 \% & 0.08 s & 42.0 \% & 42.1 \% & 0.34 s & 56.5 \% & 56.5 \% & 2.03 s & 7.2 \% & 7.2 \% & 0.11 s\\
\hline
SA & 26.9 \% & 43.1 \% & 0.16 s & 41.0 \% & 59.1 \% & 0.86 s & 43.0 \% & 58.6 \% & 0.76 s & 32.3 \% & 50.0 \% & 0.10 s\\
\hline
ISA & 40.1 \% & 46.9 \% & 2.39 s & 57.8 \% & 66.0 \% & 13.1 s & 59.3 \% & 63.1 \% & 13.6 s & 48.5 \% & 50.4 \% & 1.44 s\\
\hline
CH & 24.8 \% & 44.0 \% & 0.13 s & 35.8 \% & 61.8 \% & 0.38 s & 39.8 \% & 59.4 \% & 0.42 s & 20.8 \% & 47.8 \% & 0.14 s\\
\hline
ICH & 43.0 \% & 47.8 \% & 2.74 s & 61.3 \% & 68.6 \% & 7.50 s & 60.6 \% & 64.9 \% & 7.82 s & 48.5 \% & 50.4 \% & 2.94 s\\
\end{tabular}
\egroup
\label{results}
\end{adjustbox}

\begin{figure*}

\begin{tikzpicture}
	\begin{axis}[
		height=5cm,
		width=0.45\linewidth,
	    label={(a) Attributes on 8 bits},
	    xlabel={Initial Number of Shares},
	    style={font=\scriptsize},
	    ylabel={Decryption Time [ms]},
	    xmin=50, xmax=250,
	    ymin=0, ymax=500,
	    xtick={50,100,150,200, 250},
	    ytick={100,200,300,400},
	    legend pos=north west,
        legend style={font=\tiny},
	    ymajorgrids=true,
	    grid style=dashed,
	]

    \addplot+[]
        coordinates { (50,78) (100,155) (150,203) (200,273) (250,341)};
        \addlegendentry{Unoptimized}

    \addplot+[]
        coordinates { (50,75.6) (100,128.1) (150,162.7) (200,123.3) (250,187)};
        \addlegendentry{HC}

    \addplot+[]
        coordinates { (50,72) (100,129) (150,168) (200,114) (250,173)};
        \addlegendentry{CH}
	\addplot+[]
	    coordinates { (50,68) (100,121) (150,136) (200,90) (250,153)};
		\addlegendentry{ISA}
	\end{axis}

\node at (2,-1.2) {(a) Decryption time};
\end{tikzpicture}
\begin{tikzpicture}
	\begin{axis}[
		height=5cm,
		width=0.45\linewidth,
	    xlabel={Initial Number of Shares},
	    style={font=\scriptsize},
	    ylabel={Key generation Time [ms]},
	    xmin=50, xmax=250,
	    ymin=0, ymax=3000,
	    xtick={50,100,150,200, 250},
	    ytick={500,1000,1500,2000, 3000},
	    legend pos=north west,
        legend style={font=\tiny},
	    ymajorgrids=true,
	    grid style=dashed,
	]

    \addplot+[]
        coordinates { (50,68) (100,139) (150,181) (200,245) (250,358)};
        \addlegendentry{Unoptimized}

    \addplot+[]
        coordinates { (50,68) (100,139) (150,181) (200,245) (250,358)};
        \addlegendentry{HC}

	\addplot+[]
        coordinates { (50,178) (100,325) (150,302) (200,245) (250,521)};
		\addlegendentry{CH}

	\addplot+[]
        coordinates { (50,638) (100,885) (150,1150) (200,933) (250,3401)};
		\addlegendentry{ISA}

	\end{axis}

\node at (2,-1.2) {(b) KeyGeneration time };
\end{tikzpicture}
\caption{ABE performance tests}
\label{fig:tests}
\end{figure*}
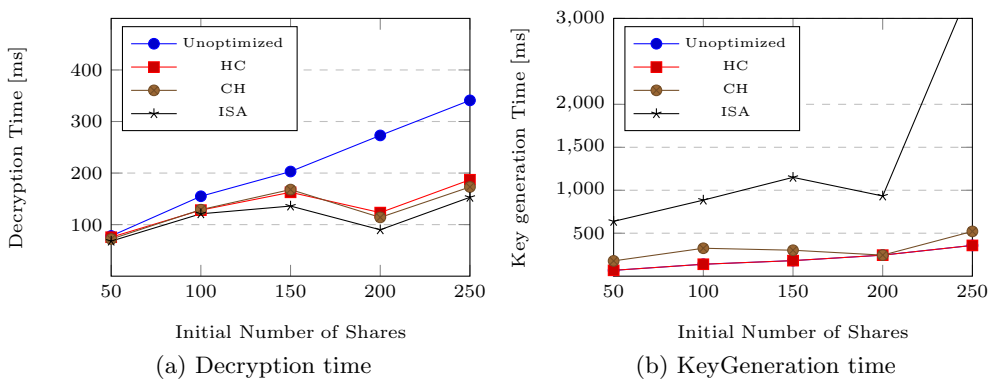

To ensure clarity and ease of interpretation of the plot in Fig. \ref{fig:tests}, we chose only three of our Heuristics for the second test, namely: HC, CH, and ISA. The first two choices were made based on the low running time while having  similar optimization results. Lastly, we have also chosen an iterated version of one of our algorithms - ISA - to be able to evaluate the trade-off between running time and optimization in the iterated versions of the algorithms.

The results can be observed in Fig.\@ \ref{fig:tests}: In a) we can see how the decryption time relates to the size of the original access structure.
 Since the Boolean circuit optimization should take place in the KeyGeneration phase, the second plot (Fig.\@ \ref{fig:tests}b) presents the relation between the increase in KeyGeneration time and the access structure size. Here, the key generation time for the unoptimized access structure is the same with the HC optimization, since the running time for HC is negligible. It is obvious that the more aggressive the optimization process is, the more time the KeyGeneration takes. Depending on the circumstances, an inefficient KeyGeneration which produces efficient Decryption algorithms may be preferred. This is actually the case in applications where the key generation takes place in some server in the cloud, while the decryption is on systems with limited capabilities, such as mobile phones, or even nodes in Wireless Sensor Networks (Such an ABE system was proposed for example in \cite{ChDa2015}).
 
 We can also observe that we have a drop in optimization for the access structure with 200 shares, compared to the one with 150 shares. This is due to the fact that we have run these tests with a single access structure for each number of shares, rather than generating multiple access structures and computing a median optimization. Also, we want to emphasize that the optimization of the Boolean circuit depends on the exact circuit. Two different circuits will have different potentials for optimization. From the two tests we ran, we can conclude that HC produces decent optimization while the computation overhead is negligible. CH and SA produce better optimization while having some small computational overhead. The iterated versions exploit the full potential of these algorithms but require a considerable additional amount of time.

\subsection{Library}

Our heuristics are publicly available on GitHub \cite{github} as a library. It provides the possibility of running a specific algorithm over a logical formula and returns the best equivalent formula found. Also, another goal of the library is to be integrated with existing ABE systems in order to optimize their performance in real-world systems.

\paragraph{Benchmarking dataset}

We have also provided \texttt{txt} files with our logical expressions which we used to test our heuristics. These can be used as references for further work on similar problems, in order to make relevant comparisons between similar algorithms with ours. The tests can be found in the \texttt{inputs} folder in our repository.

\section{Conclusions}
We have proposed multiple heuristic optimizations for minimizing monotone Boolean circuits, which, as we can see from the tests we ran, provide a substantial improvement in the circuit size. Each of our heuristics behaves differently in terms of optimization and running time, depending on the size and structure of the Boolean circuit. Since we drew our motivation from the problem of optimizing ABE systems for Boolean circuits, we emphasize that our optimizations translate into much smaller decryption keys and decryption times for these encryption systems. This can have an important impact on cloud systems that use ABE schemes to implement cryptographic access control over data: the heuristics will be applied only once when the decryption key is generated, and then, each time the respective decryption key will be used, the decryption time will be smaller compared to the scenario where we do not optimize it using one of our algorithms. For example, using the HC version of our heuristics will add almost no additional time overhead in the key generation process, while still providing an optimization between 7\% and 40\% in decryption time. We have compiled our work into a library \cite{github} written in C++, which is publicly available for anyone to use. This could be easily integrated with an existing ABE implementation written in C++, such as OpenABE \cite{openABE}.

Furthermore, there is still space for even bigger improvements, by combining the heuristic with cryptographic improvements: At the moment, our algorithms are limited by searching for Boolean circuits that are equivalent to the original one. However, by using cryptographic strategies for improving secret sharing for parts of a circuit, we could offer our heuristics more space for searching for better solutions. This remains an interesting problem to be studied.

\vspace{-0.8em}
\bibliography{abe, heuristics}
\vspace{-0.8em}
\bibliographystyle{alpha}



\clearpage

\vspace{10cm}

\end{document}